# Fiber to the Room: Key Technologies, Challenges, and Future Prospects

Jinhan Cai, *Student Member*, *IEEE*, Xiaolong Zhang, Xiang Wang, Tianhai Chang, Gangxiang Shen, *Senior Member*, *IEEE/Fellow*, *OSA*

*Abstract*—Fiber to the Room (FTTR) is a next-generation access network designed to deliver high bandwidth, low latency, and room-level optical coverage. This paper presents a comprehensive analysis of the FTTR system architecture and protocol stack, focusing on three key technical aspects: centralized scheduling and control, integrated management and maintenance, and green energy-saving mechanisms. A simplified FTTR architecture based on the convergence of the medium access control (MAC) and physical (PHY) layers is introduced to enhance coordination and scheduling efficiency. An extended remote management scheme, based on the optical network unit management and control interface (OMCI), is described to enable unified control across main fiber units (MFUs) and sub-fiber units (SFUs). Furthermore, a service-aware energy-saving framework is discussed for dynamic power optimization. The paper also explores the integration of artificial intelligence (AI) and passive sensing into FTTR systems to support intelligent scheduling, energy management, and environment-aware optimization. These insights provide technical guidance for the scalable deployment and future evolution of FTTR networks.

## I. INTRODUCTION

WITH the continuous upgrading of household broadband, users' expectations for network experience are evolving from basic availability to guaranteed performance. For example, new video services offering 4K/8K ultra-high-definition videos require networks capable of delivering hundreds of Mb/s of bandwidth [1]. In addition, new services such as online education, streaming media, Augmented Reality (AR) and Virtual Reality (VR) are placing high demands on mobility, network latency, packet loss rate, and video quality [2]. On the other hand, in enterprise and industrial settings, smart offices and intelligent manufacturing systems demand higher stability and reliability from access networks [3]. To address these challenges, the European Telecommunications Standards Institute (ETSI) has been defining and standardizing the Fifth Generation Fixed Network (F5G) since 2020 [4], [5]. The F5G's effort aims to achieve ubiquitous fiber deployment and support three critical features: enhanced Fixed BroadBand (eFBB), Full Fiber Connectivity (FFC), and Guaranteed Reliable Experience (GRE) [4], [5].

Passive Optical Network (PON) and Wi-Fi are critical technologies supporting F5G access networks. PON is widely recognized as a leading broadband access technology [6]. The International Telecommunication Union - Telecommunication Standardization Sector (ITU-T) has identified 50 Gb/s single-wavelength PON as the next-generation PON technology, following XG(S)-PON [7], [8]. Wi-Fi, known for its high speed, flexibility, and convenience, has become the predominant technology for indoor wireless communication. The 802.11ax (Wi-Fi 6) standard [9], which supports a maximum data rate of 9.6 Gb/s, has seen wide adoption. The latest 802.11be (Wi-Fi 7) standard [10] has been released, and the development of 802.11bn (Wi-Fi 8) [11] is currently underway.

However, in current Fiber to the Home (FTTH) architectures that rely on Wi-Fi for home access networks [12], ensuring quality of service (QoS) remains a challenge—even with the latest Wi-Fi technologies. This is primarily due to several limitations: (1) home Wi-Fi networks utilize copper and air interfaces, which inherently limit capacity; (2) interference among adjacent Wi-Fi access points (APs) is common in residential environments [13]; and (3) the random-access nature of Wi-Fi cannot guarantee the low latency required by advanced applications [14]. Therefore, enhancing Wi-Fi throughput and QoS is essential for improving the overall performance of home access networks.

To address the aforementioned issues, ITU-T, the China Communications Standards Association (CCSA), and the ETSI F5G Industry Specification Group have jointly proposed a reference architecture for fiber in-premises (FIP) networks, known as Fiber to the Room (FTTR) [5]. Building upon the FTTH architecture, this design further extends the deployment of optical fiber to individual rooms to improve the performance of indoor networks. Consequently, FTTR promises to enable gigabit-level access throughout the households. In the context of international standardization, the ETSI F5G Industry Specification Group released a use case report in 2020, which, for the first time, defined FTTR for home networks [15]. In the same year, the ITU-T SG15 Q3 group published a series of technical reports [16], supplementary documents and

This work was supported in part by the National Nature Science Foundation of China under Grant W2411058. (Corresponding author: Gangxiang Shen; e-mail: shengx@suda.edu.cn).

Jinhan Cai and Gangxiang Shen are with Jiangsu Provincial Laboratory for Advanced Optical Transmission and Switching Networks, Jiangsu New Optical Fiber Technology and Communication Network Engineering Research Center, Suzhou Key Laboratory of Advanced Optical Communication Network Technology, and School of Electronic and Information Engineering, Soochow University, Jiangsu Province 215006, P. R. China. Xiaolong Zhang, Xiang Wang and Tianhai Chang are with Huawei Technology, P. R. China.

recommendations [17], focusing on the requirements for deploying FTTR in small and medium-sized enterprise environments. In December 2023, ITU-T finalized the recommendation for the FTTR system architecture and officially designated the FTTR specification series as the fiber in-premises network (FIN) [18]. This series is formally identified by the prefix G.fin, aligning it with ITU-T's G-series recommendations. The recommendation specifies several key technical requirements, including a reference model for the centralized coordination network architecture, physical layer requirements, system-level requirements, and system management functions. In July 2024, ITU-T released the physical layer specification [19] and the data link layer specification [20], further enhancing the standardization of FTTR.

In China, CCSA completed a research report on FTTR in 2020, emphasizing that fiber-based access networks represent the most promising infrastructure for indoor communications [21]. In May of the same year, Huawei and China Telecom jointly released an FTTR white paper [22], which explored innovative FTTR-based services and corresponding technical solutions. In 2022, the Broadband Development Alliance published a revised version of the FTTR white paper [23], introducing FTTR-related policies and developments across the industry chain. It also provided a preliminary analysis of technical features, case studies, business models, and the evolution of the industry ecosystem. In January 2024, under the promotion of China Mobile, ITU-T officially initiated the project "Coordinated Management of Access and Fiber In-Premises Networks (G.sup.CMAFP)" [24]. This standard aims to establish a coordinated management architecture between PON and FTTR. The technical specifications are currently under development and are being aligned with related ITU-T standards.

This paper investigates the key technical challenges and potential applications of FTTR, with the aim of providing insights into its future development. The remainder of the paper is organized as follows: Section II presents an overview of the FTTR system based on the current G.fin standards. Section III examines the technical challenges within the FTTR system, analyzes potential issues along with feasible solutions, and outlines possible directions for future advancement. Section IV discusses key technologies that can be integrated into the FTTR framework and explores their applications. Finally, Section V concludes the paper.

## II. FTTR SYSTEM OVERVIEW

This section provides an overview of the FTTR system as defined in the current G.fin standard. It outlines the overall network architecture, describes the functional and protocol models, and examines the integration of wireless technologies within the FTTR framework.

### A. Network Architecture

Fig. 1(a) illustrates the architecture of a G.fin-based fiber in-premises (FIP) network. The architecture consists of two parts: an FTTH segment and a FIP network. The FTTH segment leverages existing optical access technologies, such as 10G-EPON or XG(S)-PON. The FIP network is further divided into the FTTR domain and the user-side wireless/wired interfaces. The FTTR domain comprises a Main Fiber Unit (MFU), an Indoor Fiber Distribution Network (IFDN), and multiple Sub Fiber Units (SFUs). The MFU connects to SFUs via the IFDN using a point-to-multipoint (P2MP) topology. The MFU replaces the conventional Optical Network Unit (ONU) to act as the data, control, and management hub in the FIP network. It functions as both the PON termination point as well as the access gateway for the in-premises network. Both MFU and SFUs support wireless (e.g., Wi-Fi) and wired (e.g., Ethernet) interfaces for user devices, such as smartphones and laptops.

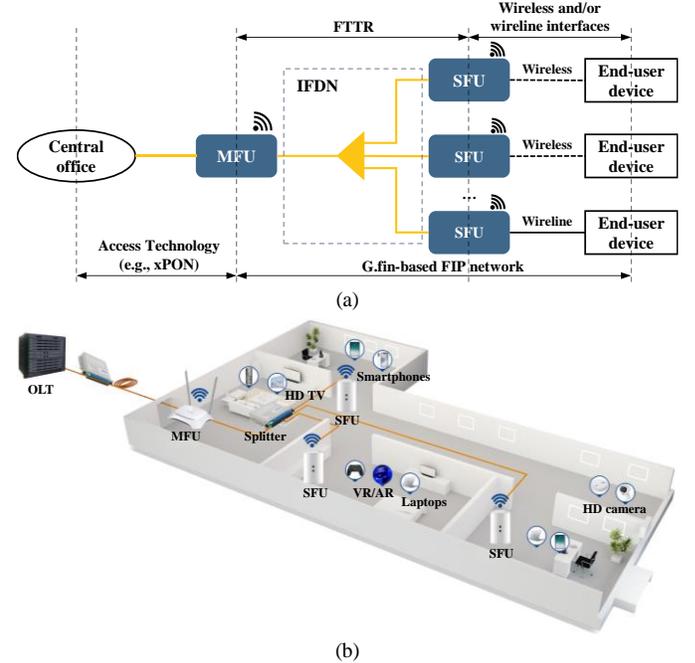

Fig. 1. (a) Architecture of a G.fin-based FIP network. (b) Example of FTTR application in a residential environment.

Fig. 1(b) illustrates an example deployment of FTTR in a residential environment. In such scenarios, devices like smartphones, HDTVs, high-definition cameras, and VR/AR headsets require stable end-to-end connectivity to deliver high-quality user experiences. FTTR improves wireless coverage and access quality by deploying dedicated SFUs in each room. With per-room deployment, power control techniques can limit signal coverage, minimizing interference between neighboring SFUs. Moreover, increased AP density reduces the number of clients per SFU, mitigating channel congestion. In this architecture, the MFU acts as the centralized controller for both FTTR and Wi-Fi components. It supports remote management of all SFUs, enables coordinated channel optimization to avoid interference, and facilitates intelligent roaming, thereby reducing latency during AP handovers.

### B. Functional and Protocol Reference Model

Fig. 2 illustrates the functional architecture of a G.fin-based FIP network. The architecture is organized into three logical planes: management plane, control plane, and data plane. In the management plane, the MFU centrally manages the entire network via the MFU management entity (MME). The MFU uses a southbound interface to transmit management messages for configuring the SFU management entity (SME), thereby enabling the management and maintenance of the G.fin data link layer (DLL) and physical layer (PHY) within each SFU. Simultaneously, the MFU reports the overall G.fin network status to the management system through a northbound interface. The management system is responsible for monitoring the carrier's assets and device states, as well as optimizing the performance of the G.fin network.

In the control plane, the MFU control entity (MCE) sends status collection requests to the SFU coordination entity (SCE), while the SCE continuously collects and reports network status back to the MCE. Based on the feedback, the MFU applies system-wide transmission strategies to coordinate the resource allocation and sends the coordination information to the wireless interfaces on the SFUs, enabling joint optimization between the optical and wireless domains.

The data plane provides the transport channels for service, management, and control data. At the MFU, downstream traffic is first processed by the physical media dependent (PMD) layer of the PON, followed by frame decapsulation at the transmission convergence (TC) layer. Ethernet frames are bridged or routed between the PON TC layer and the G.fin DLL via the L2+ layer. G.fin DLL encapsulates both user payload and control/management messages into DLL frames, which are further processed by the G.fin PHY and transmitted over the IFDN to the SFUs. On the SFU side, Ethernet frames are bridged between the G.fin DLL and the Ethernet/Wi-Fi medium access control (MAC) layer, and are finally transmitted to end-user devices through the wired or wireless interfaces provided by the SFU.

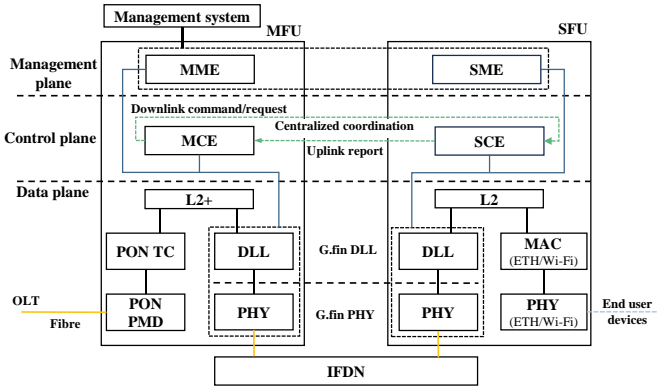

Fig. 2. Functional architecture of a G.fin-based FIP network.

Figure 3 illustrates the protocol stack of the G.fin system. As shown in Fig. 3(a), the data link layer (DLL) of G.fin comprises three sublayers: Application Protocol Convergence (APC), Logical Link Control (LLC), and Media Access Control (MAC). During transmission, each service data unit (SDU) entering the APC sublayer is mapped to an application protocol data unit (APDU) using the FIN encapsulation method (FEM). Based on this mapping rule, each SDU destined for a target node is encapsulated into an APDU. The APC sublayer also identifies classification tags (e.g., priority labels), and assigns each APDU to one or more transmission queues according to its classification. These classification indicators are then used by the LLC sublayer to ensure that the QoS requirements of the corresponding services are satisfied. The APDUs are subsequently delivered to the LLC sublayer, which also receives management and control data from the DLL management and control entity. These control messages are formatted as fiber management and control interface (FMCI) data units (FMCI-DUs) and WLAN management and control interface (WMCI) data units (WMCI-DUs), which are used for managing the optical and wireless links, respectively. In the LLC sublayer, the APDUs, FMCI-DUs, and WMCI-DUs are encapsulated into FEM frames. FEM headers are added to the resulting frames before they are passed to the MAC sublayer. The MAC sublayer concatenates the FEM frames into MAC protocol data units (MPDUs) and forwards them to the PHY layer according to the scheduling strategy specified by the LLC (e.g., multi-queue scheduling). Additionally, media access rules (such as port ID mapping, packet filtering, and dynamic time allocation) are configured by the DLL management and control entity. During reception, MPDUs received by the MAC sublayer are parsed into FEM frames. The LLC sublayer extracts the original APDU, FMCI-DU, and WMCI-DU from these frames and forwards them to the APC sublayer and the DLL management and control entity, respectively. The APC sublayer then reconstructs the SDU from the APDU and delivers it to the application layer.

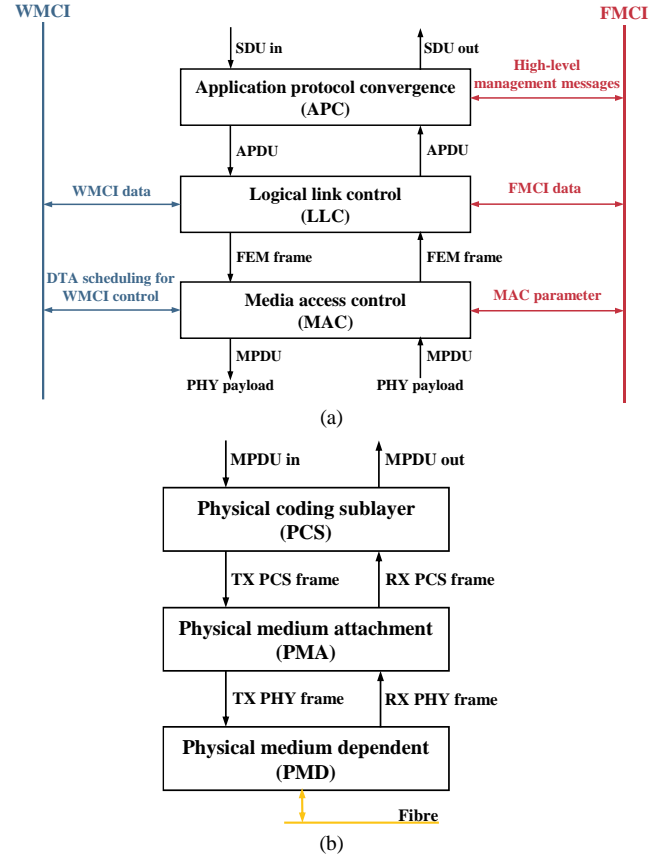

Fig. 3. Functional protocol stack of the G.fin system: (a) the layered structure of DLL; (b) the layered structure of PHY.

As shown in Fig. 3(b), the G.fin physical layer comprises three sublayers: physical coding sublayer (PCS), physical medium attachment (PMA), and physical medium dependent (PMD). In the transmission direction, the MPDU is mapped to the payload of a PCS frame, and a PCS header is appended. The header includes physical layer operation, administration, and maintenance (PLOAM) messages, as well as a time assignment map (TAmap), which supports dynamic bandwidth allocation. The PCS frame is then passed to the PMA sublayer for forward error correction (FEC) encoding and scrambling. The encoded frame is then modulated and transmitted over optical fiber by the PMD sublayer. Upstream and downstream wavelength division multiplexing (WDM) is implemented at the PMD sublayer. In the receiving direction, the PMD sublayer converts an incoming optical signal into an electrical signal and forwards it to the PMA sublayer. The PMA sublayer performs FEC decoding and descrambling on the received PHY frame. The PCS sublayer processes the decoded frame header and payload, recovers the original MPDU, and passes it to the data link layer. In parallel, control information parsed from the PCS header is forwarded to the PHY layer management entity for further processing.

*C. Wireless Access Technologies for FTTR*

In FTTR networks, the SFU provides the last ten meters of user access through wireless technologies. Wi-Fi has been recommended as the primary wireless access technology in FTTR standards due to its broad applicability and backward compatibility [25]. With the evolution of Wi-Fi technology, the theoretical peak downlink rate has reached 9.6 Gb/s with Wi-Fi 6 and up to 30 Gb/s with Wi-Fi 7 [26]. The advancement of Wi-Fi is not limited to improvements in data rates, but also includes significant functional and architectural enhancements. Wi-Fi 6 introduced orthogonal frequency-division multiple access (OFDMA) in both uplink and downlink transmissions [27]. Wi-Fi 7 supports multi-link operation (MLO), enabling devices to function as multi-link devices (MLDs) [28]. Wi-Fi 8 is designed with coordinated transmission across multiple access points (APs) as a key objective [11], [29]. This progression—from resource-level coordination to multi-link and multi-device coordination—closely aligns with the system objectives of FTTR.

In addition to the Sub-6 GHz spectrum, millimeter-wave (mmWave) communication has been considered effective for high-data-rate transmission applications due to its wide available bandwidth [30]. To support indoor mmWave communications, a series of standards have been published by the IEEE, including IEEE 802.11ad [31] and the more recent IEEE 802.11ay [32] under the 802.11 working group, as well as IEEE 802.15.3c [33] under the 802.15 working group. In China, the State Radio Regulatory Commission (SRRC) authorized the use of the 42–48 GHz frequency band (Q-band) in 2013. Based on this allocation, the corresponding IEEE 802.11aj/Q-LINKPAN transmission standard was developed by researchers at Southeast University, China [34], [35]. However, indoor mmWave communication is constrained by significant wall penetration loss, limited coverage, and high backhaul bandwidth requirements, which have hindered its widespread deployment in recent years. Within the FTTR architecture, the deployment of SFUs in individual rooms can effectively enhance mmWave coverage in indoor environments. The physical separation between rooms also helps reduce inter-room interference among signals from different APs.

Additionally, the optical fiber links between the MFU and SFUs provide sufficient backhaul capacity to meet the high-data-rate demands of mmWave communication. Existing studies have explored the feasibility of Q-band mmWave in FTTR systems from various perspectives, including signal transmission [36], MAC layer protocols [37], circuit-level design [38], and sensing applications [39]. Future access solutions may also incorporate emerging technologies such as visible light communication (VLC) and optical wireless communication (OWC) [40]-[42], contributing to the realization of fully optical connectivity.

### III. ENABLING TECHNOLOGIES FOR FTTR NETWORKS

This section reviews the research progress, feasible solutions, and potential challenges in the following three key areas: centralized scheduling and control, integrated management and maintenance, and green energy-saving technologies. First, to reduce system latency and jitter, FTTR requires centralized control at the MFU, as well as optimized air interface resource allocation and enhanced coordination between optical and wireless links. Second, large-scale FTTR deployment increases the demand for efficient network management and maintenance. Network operations must be further optimized to enhance efficiency and reduce operational costs. Finally, with the growing adoption of FTTR, energy consumption has become a significant concern. As a result, green and energy-efficient technologies have emerged as a key focus in FTTR research.

*A. Centralized Scheduling and Control*

FTTR is designed to deliver indoor network services with guaranteed QoS for end users, requiring coordinated resource management and channel access across optical and wireless links. As illustrated in Fig. 4, a centralized coordination architecture based on the G.fin framework enables this functionality. The controller continuously gathers information on transmission demands and traffic characteristics (e.g., buffer occupancy, service priority, latency requirements), as well as network environment parameters (e.g., channel conditions between the MFU and SFUs, interference levels). Based on this information, it formulates control strategies to coordinate various network components, supporting functions such as global handover and real-time link monitoring. According to the QoS requirements of each service, the controller issues instructions to the SFU modules, including the Wi-Fi unit, Ethernet switch, and buffer. After receiving the control message, the SFU reports local network status back to the MFU. Additionally, the controller sends dynamic commands to the local coordinator in each SFU, which is equipped with integrated scheduling capabilities for both fiber and wireless links—such as optimizing air interface time, frequency, and spatial resource allocation, or scheduling fiber link transmissions—to enhance overall resource utilization efficiency.

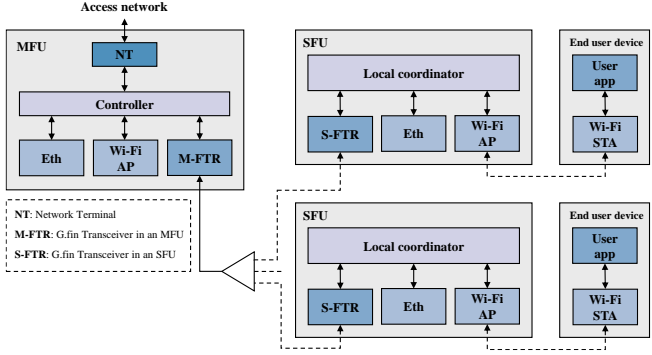

Fig. 4. Centralized coordination network architecture based on G.fin.

Based on the G.fin standard, Huawei has proposed the Centralized Wireless Access Network (C-WAN) architecture as an engineering solution to implement centralized scheduling in FTTR scenarios [43]. During operation, the SFU uses its local processor to collect status information from both the optical link and the Wi-Fi interface, and reports this information to the MFU via the Network Control Interface (NCI). The controller then performs downlink scheduling based on the global network status and sends control instructions through the NCI, enabling the SFU processor to manage the local Wi-Fi module accordingly. This architecture allows the MFU to dynamically allocate air interface transmission opportunities for each SFU in every downlink cycle [14].

Under the centralized control architecture, Wi-Fi-related functions are migrated from distributed processing to centralized scheduling. For example, during downlink transmission, when multiple SFUs operate within overlapping wireless coverage areas, interference and channel contention may occur. In such cases, each SFU reports its MAC buffer status and service priority levels to the MFU, which then coordinates their channel access behavior. Specifically, the MFU schedules SFUs experiencing co-channel interference based on their respective traffic volumes and service priorities. Control messages are sent via the downlink optical link, instructing certain SFUs to temporarily defer channel contention while granting others access. Once the granted SFU completes its transmission, the MFU authorizes the next SFU to transmit data, thereby enabling contention-free utilization of the air interface.

In the uplink direction, centralized control can be integrated with the OFDMA mechanism introduced in Wi-Fi 6, where the AP allocates resource units (RUs) to individual STAs via trigger frames to support contention-free uplink transmissions [44]. However, due to the limited number of RUs, precise scheduling is critical to avoid resource conflicts and inefficiencies. With centralized coordination, the MFU instructs SFUs to initiate OFDMA-based transmissions in a sequential manner. Upon sending a trigger frame, an SFU estimates the total uplink data volume based on the RU allocation and requests corresponding optical uplink bandwidth from the MFU. Once the data is received via OFDMA at the SFU, it can be directly forwarded to the MFU, thereby minimizing queuing delays and improving overall transmission efficiency.

In the current FTTR architecture, each SFU integrates a complete Wi-Fi module with independent access and scheduling functionalities, and communicates with the MFU solely through the G.fin protocol. Moreover, the optical link and the Wi-Fi system operate independently at the physical level, with separate MAC and PHY layers that interact only via control plane signaling. Under such an architecture, the centralized control capability of the MFU is constrained, making it challenging to support fine-grained management functionalities such as frame aggregation, access parameter tuning, roaming timing coordination, and target AP selection. To overcome these limitations, we propose a decoupled design of the Wi-Fi protocol stack at the SFU, in which selected control and protocol processing functions are migrated to the MFU. This architecture enables centralized management and resource pooling across multiple SFUs, thereby enhancing network coordination and scheduling efficiency.

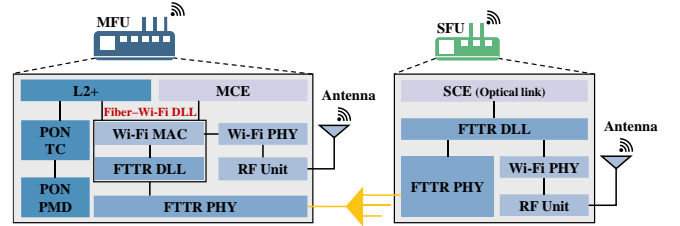

Fig. 5. Architecture based on the integration of DLL and Wi-Fi MAC.

**MAC Layer-based Integration:** Fig. 5 illustrates the FTTR architecture based on the integration of the data link layer (DLL) with the Wi-Fi MAC. In this design, the DLL module at the MFU, which was originally decoupled from the Wi-Fi stack, is enhanced to incorporate Wi-Fi MAC functionality, forming a unified Fiber-Wi-Fi DLL. As a result, the Wi-Fi module is simplified, retaining only the PHY layer and radio frequency (RF) unit. The SFU comprises the FTTR PHY, FTTR DLL, a local control entity, and a lightweight Wi-Fi PHY module. With MAC functionality centralized at the MFU, the SFU's control entity is responsible solely for optical link-related control. In the downlink direction, Ethernet packets first enter the Fiber-Wi-Fi DLL at the MFU, where the Wi-Fi MAC sublayer encapsulates them into MPDUs. These MPDUs are then transmitted over the air interface via the Wi-Fi PHY and RF unit. For data destined for an SFU, the FTTR DLL processes the packets, and the control entity inserts necessary optical control fields before forwarding them over the optical link. Upon receiving the downlink frame, the SFU extracts the optical control information via the FTTR DLL and delivers it to the local control entity. The MPDU is subsequently processed by the local Wi-Fi PHY and RF unit for air interface transmission. The uplink process follows the reverse path.

In this architecture, the optical link carries Wi-Fi MAC frames instead of Ethernet frames, and Wi-Fi MAC functionality is centralized at the MFU and integrated with the optical link at the protocol level. Compared with the conventional architecture, in which the FTTR DLL and the Wi-Fi stack are functionally decoupled, the MFU was previously limited to managing only high-level behaviors of the SFU, such as channel contention and transmission initiation, through control messages. Under the proposed design, the MFU obtains fine-grained control over frame encapsulation and transmission timing. In OFDMA-based scenarios, it can further configure the RUs assigned to each STA within the trigger frame. Consequently, user management, queue scheduling, frame construction, and resource allocation can all be centralized at the MFU, enabling more precise and coordinated Wi-Fi scheduling across the network. Moreover, centralization enhances the system's capacity for global resource optimization and the application of machine learning, as the MFU has a holistic view of the network environment across multiple SFUs. This enables coordinated scheduling to minimize cross-link interference and maximize spectral efficiency, while also supporting continuous learning from historical data to predict traffic patterns, optimize transmission strategies, and enable proactive network adjustments.

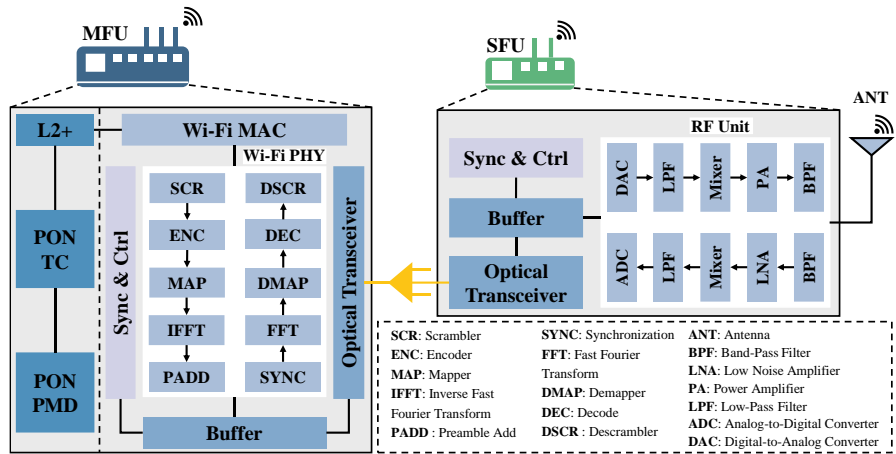

Fig. 6. FTTR architecture based on the integration of the MAC and partial PHY of the MFU and SFU.

**PHY Layer-based Integration:** Fig. 6 illustrates an FTTR architecture based on the integration of the MAC and partial PHY functionalities between the MFU and SFU. In this design, the SFU is simplified from a fully functional data-processing device into a relay node responsible solely for converting wireless signals into optical signals. It integrates an RF unit, a synchronization and control module, an optical transceiver, and a buffer unit. The RF unit handles the conversion between wireless RF signals and digital baseband signals. The synchronization and control module performs three main functions: (1) ensuring clock synchronization between the SFU and the MFU; (2) estimating the required uplink bandwidth and parsing bandwidth allocation instructions from the MFU; and (3) controlling the SFU's uplink and downlink transmission. The optical transceiver enables signal conversion between electrical and optical domains. After simplification, the SFU can be regarded as a remote RF unit extended from the MFU via optical fiber, while the MFU functions as a central node responsible for user access control and data transmission. Consequently, the DLL and PHY functions originally defined by the G.fin standard for MFU-SFU interaction can be significantly streamlined. Besides PON-related modules, the MFU integrates the migrated Wi-Fi module, a synchronization and control module, and an optical transceiver. The Wi-Fi module implements a complete set of MAC and PHY functionalities. The synchronization and control module is responsible for exchanging control messages with its counterpart at the SFU. Its primary functions include: (1) ensuring clock synchronization between the MFU and each SFU; (2) receiving uplink bandwidth requests from SFUs and estimating the corresponding allocated bandwidth; (3) controlling downlink data transmission from the MFU.

In the downlink direction, the Wi-Fi MAC layer at the MFU receives packets from the bridging layer (e.g., Ethernet packets) and encapsulates them into MAC frame payloads. The MAC layer also performs resource coordination and scheduling, embedding control information in the MAC header. The header and payload together form the MPDU, which is passed to the Wi-Fi PHY layer. The PHY layer performs scrambling, encoding, modulation, subcarrier mapping, and IFFT to generate the OFDM-modulated signal. After preamble addition, a complete PPDU is formed and the resulting digital baseband signal is stored in the buffer. Based on the bandwidth requests reported by SFUs and the current downlink traffic status, the synchronization and control module determines bandwidth allocation and scheduling timing. The downlink baseband signal and SFU scheduling information are then transmitted over the optical link under the control of the synchronization and control module at MFU. At the SFU, the optical transceiver converts the received optical signals into electrical signals. The synchronization and control module parses the embedded bandwidth allocation instructions, and the corresponding digital baseband signal is buffered. The RF unit then processes the buffered signal through digital-to-analog conversion (DAC), low-pass filtering, frequency up-conversion via a mixer, power amplification, and band-pass filtering. The resulting RF signal is transmitted over the air interface via the antenna.

In the uplink direction, the SFU receives RF signals from an STA. These signals are processed by standard front-end modules, including band-pass filtering (BPF), amplification (LNA), frequency down-conversion (Mixer), low-pass filtering (LPF), and digitization (ADC). The resulting digital baseband signal is temporarily stored in a buffer. During this process, the synchronization and control module measures the sampling duration and estimates the required uplink time slot length based on hardware parameters such as the sampling rate, bit width, and optical link rate. Within the uplink time slot assigned by the MFU, the SFU transmits both the baseband signal and the corresponding bandwidth request to the MFU over the optical link. At the MFU, the optical transceiver receives the uplink signal, converts it into an electrical format, and temporarily stores it in the buffer. The synchronization and control module then parses the embedded bandwidth request information for subsequent scheduling decisions. The digital baseband signal is processed by the Wi-Fi PHY layer, including synchronization, FFT, subcarrier demapping, decoding, and descrambling, to recover the MPDU. The Wi-Fi MAC layer parses and verifies the MPDU. The Wi-Fi MAC layer parses and verifies the MPDU. Control messages are processed locally at the MAC layer, while regular data frames are delivered to the bridging module for forwarding or switching.

In this architecture, all wireless processing capabilities are centralized at the MFU, with Wi-Fi digital baseband signals transmitted over the optical link. This design decouples RF access from baseband processing, greatly simplifying both the design and power requirements of the SFU. By consolidating signal processing functions at the MFU, the system can extract fine-grained physical-layer features such as channel state information, interference patterns, and link quality metrics. These high-resolution features provide a strong foundation for enabling AI-defined radio capabilities, including intelligent interference prediction, dynamic beamforming and tracking, and adaptive modulation and coding scheme (MCS) selection [45], [46].

*B. Integrated Management and Maintenance*

Compared with FTTH, FTTR imposes higher demands on the management and control system, as it must support functions such as device discovery, status detection, configuration provisioning, and fault localization across both MFUs and SFUs [23]. The core management and control capabilities of the FTTR system include: (1) network visualization, which displays the network topology, device status, and key performance indicators such as Wi-Fi interference; (2) configuration management, enabling remote setup of authentication methods, VLAN segmentation, and QoS policies for MFUs and SFUs; (3) performance monitoring and data collection, allowing continuous evaluation of link performance across wireless interfaces and LAN ports; and (4) fault detection and alarm mechanisms, which identify device failures or link disruptions and enhance fault response efficiency. To support these functions, the system must offer remote and integrated management across the entire communication channel from the OLT to each SFU.

For remote management of MFUs, G.fin defines two primary approaches. The first is a PON-compatible method based on the ONU Management and Control Interface (OMCI) [47], in which the OLT manages MFUs as ONUs for purposes such as registration, configuration, and status reporting. The second is an IP-based approach (e.g., Broadband Forum TR-069 [48] or TR-369 [49]), where an IP protocol agent is inserted between the PON TC layer and the MME of the MFU, enabling the remote management system to manage the device directly without involving the OLT.

For remote management and control within the FTTR domain, two possible solutions exist. The first is a segmented management approach, where the OLT manages the MFU via OMCI, and the MFU acts as an intermediate node providing OMCI-like or IP-based management services to downstream devices, resulting in two independent management hierarchies and imposing stringent requirements on the MFU in terms of protocol parsing and message forwarding [18]. A more advanced alternative involves constructing an integrated management framework, where the OLT manages both MFUs and SFUs through a centralized OMCI path. In this approach, SFUs are treated as logical extensions of ONUs, inheriting formats and processes defined within the existing OMCI framework, which reduces the functional complexity of MFUs and enhances the OLT's ability to centrally manage the home network. Furthermore, as outlined in the ITU-T G.fin DLL recommendations, the FTTR system shares a high degree of structural similarity with the TC layer in PON [50], providing a standardized basis for reusing OMCI message formats and management entities within FTTR systems. Therefore, extending OMCI to include MFU and SFU management is a viable approach for enabling unified remote control across both the access and in-premise network domains.

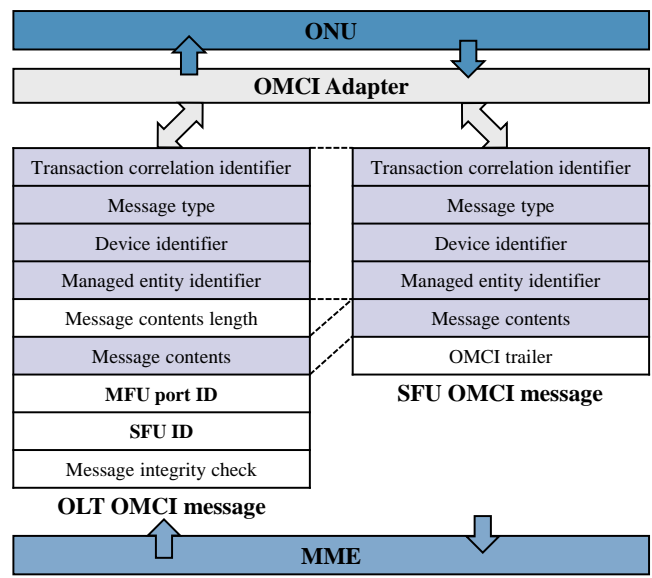

Fig. 7. Integrated remote management scheme for FTTR using the extended OMCI protocol.

Fig. 7 illustrates an integrated remote management scheme for FTTR based on the extended OMCI protocol. In this scheme, the OMCI message format is unified across both the FTTH and FTTR networks. By introducing FTTR segments into the original OMCI message channel, unified management and forwarding of cross-domain messages are achieved. For this purpose, an OMCI adapter is introduced above the MME within the MFU to enable OMCI message conversion between the OLT and the two management regions within the FTTR. Given the large number of SFUs, all SFUs under the same MFU share a single port for OMCI message transmission, conserving port resources. The specific forwarding path is determined by the expansion field within the OMCI message. According to ITU-T G.988 [47], the extended OMCI format reserves the first 8 bytes (for essential management headers) and the last 4 bytes (for message integrity checks), while allowing flexible extensions in the content section. The message content length field specifies the actual length of the message. Based on this, two additional bytes are appended at the end of the content field to identify the MFU port ID and SFU ID, respectively. When the OLT sends an extended OMCI message, the OMCI adapter converts it to the standard OMCI format and uses the embedded MFU port ID and SFU ID to route the message precisely to the target SFU. Conversely, when an SFU reports a standard OMCI message, the adapter appends the port ID and SFU ID fields to enable the OLT to identify the source of the message. OMCI messages are transmitted as data payloads over transmission containers (T-CONTs), which share the upstream bandwidth resources. To prevent management traffic from competing with regular service data for network bandwidth, it is recommended to allocate a dedicated T-CONT specifically for OMCI messages.

TABLE I
CLASSIFICATION AND COMPARISON OF TYPICAL ENERGY-SAVING STRATEGIES IN FTTR SCENARIOS

| Strategy | Condition | Complexity | Potential | Note |
|---|---|---|---|---|
| RF Off | IoT, low activity | Low | Medium | Fast response |
| TX Power Adjustment | Moderate load, concentrated access | Low | Low | Adaptive control |
| SFU Light Sleep | Short-term idle | Medium | Medium | Fast wake-up |
| SFU Deep Sleep | Long-term idle | High | High | Optical-wireless coordination |
| Optical Rate Adaptation | Optical idle time | Medium | Medium | PON-standard extension |
| Global Policy Switching | Traffic/user prediction | High | High | AI-driven optimization |

*C. Energy-Saving Technologies*

Compared to traditional access architectures, FTTR systems involve a significantly larger number of devices. To date, systematic studies on the energy efficiency of FTTR networks remain limited. Existing results indicate that, in the absence of energy-saving mechanisms, the average daily energy consumption of an FTTR deployment is approximately 1.5 times higher than that of a conventional FTTH + Wi-Fi architecture, using a typical four-room household as a reference scenario [51]. The FTTR network adopts a centralized control architecture, in which MFUs are equipped with global sensing and scheduling capabilities to support remote monitoring and power management based on the operational status of SFUs. Leveraging this architecture, we propose a sleep-mode power-saving mechanism for SFUs. Specifically, when no user activity is detected for an extended period, an SFU enters a light sleep state and reports its status to the MFU. Once the MFU confirms that all SFUs are in sleep mode, it issues control commands to transition them into deep sleep. In this state, each SFU maintains only essential functions and periodically activates its receiving module to detect control signals from the MFU. During this period, the MFU continues normal operation to ensure basic Wi-Fi coverage. Upon detecting new user activity, the MFU promptly wakes the SFUs into a low-power idle state, and then gradually transitions them to active states based on actual usage demands.

This subsection presents the FTTR energy-saving process from a system-level perspective, which is divided into three steps: *state data acquisition*, *service scenario identification*, and *dynamic policy selection*. First, the MFU collects multi-dimensional state data in real time via management and control channels. These data include Wi-Fi and optical interface traffic, device access status, received signal strength indicator (RSSI), and more [52]. Second, the collected features are processed by local service scenario recognition models to identify the current service type (e.g., short video, online gaming) and traffic load level (e.g., background or bursty traffic) [53]. Finally, based on the recognition results, the MFU selects an appropriate energy-saving policy and dispatches control instructions to the corresponding SFU to execute module-level energy-saving operations, such as RF scheduling, power adjustment, and band deactivation [54].

For scenario identification, two critical dimensions should be considered: (1) service awareness: By analyzing traffic types and load patterns, the system can anticipate potential user behavior and proactively trigger early activation or timely exit from energy-saving modes. For instance, if sudden traffic surges or latency-sensitive interactive services are predicted, the corresponding SFUs should be immediately switched to the active state to maintain service continuity. (2) user behavior detection: User mobility can lead to roaming, resulting in AP handovers. These handovers not only affect the energy state of the current SFU but also initiate the configuration of the target SFU. By accurately identifying user activity, the system can optimally trigger state transitions, reduce switching delays, and improve responsiveness.

In terms of energy-saving strategy selection, different approaches should be tailored to the specific device types and functional modules involved. For scenarios involving resident IoT devices, SFUs should avoid entering full sleep states and instead adopt dynamic power-saving modes, such as partial RF channel deactivation, transmission power reduction, or frequency band restriction. These measures help maintain continuous connectivity while effectively reducing energy consumption. When transitioning to sleep states, coordination between the optical and wireless modules becomes essential. Due to differences in wake-up latency and control protocols between these modules, a synchronization mechanism is required to ensure smooth transitions between sleep and active states. In addition, buffering and aggregation strategies must be implemented to manage data accumulated during sleep periods, thereby avoiding packet loss or excessive transmission delays caused by state transitions. Table I summarizes typical energy-saving strategies in FTTR scenarios.

Looking ahead, the application of AI technologies to energy conservation decision-making in FTTR networks holds significant potential. By leveraging data from multiple sources, the MFU can perform adaptive optimization to enhance the accuracy and responsiveness of strategy selection. Fundamentally, energy saving in FTTR is a data-driven, intelligence-enabled, and orchestrated system-level endeavor. The key challenge lies not only in reducing overall energy consumption but also in achieving dynamic, fine-grained, and highly reliable energy management without compromising service quality. Addressing this challenge will remain a critical focus of future FTTR research.

## IV. FTTR+X: INTEGRATED INTELLIGENCE AND SENSING

FTTR establishes an in-premise optical network infrastructure with centralized control, providing a reliable and efficient platform for the deployment of diverse technologies and the enablement of emerging applications. At the same time, the integration of new technologies and services can further optimize FTTR network performance, creating a positive feedback loop of mutual enhancement.

### A. AI-Empowered FTTR Networks

AI, as a key driver in the evolution of communication systems, is increasingly being integrated into FTTR networks. It enables FTTR to achieve higher levels of intelligence in resource scheduling, energy efficiency optimization, and fault management.

In resource scheduling, AI can leverage techniques such as traffic prediction [55], flow identification and classification [56], and multi-agent reinforcement learning (MARL) to support dynamic and optimized allocation of critical resources, including bandwidth, time slots, and channel access opportunities [57]. For example, time-series modeling can be used to predict service demand trends for different terminals, enabling proactive resource allocation to mitigate potential congestion. Real-time traffic identification and classification can differentiate between service types (e.g., video conferencing, file downloads, IoT data uploads), enabling the application of differentiated scheduling strategies. Latency-sensitive services such as video conferencing can be prioritized in terms of bandwidth and delay guarantees, while non-time-sensitive services can be deferred to idle periods or assigned to lower-priority channels. Furthermore, through MARL, each terminal can be modeled as an autonomous agent that dynamically adjusts parameters [58] such as backoff window size, clear channel assessment (CCA) thresholds, and transmission time slots. This distributed decision-making approach improves overall network throughput and reduces latency. Compared to traditional static or rule-based strategies, AI-based scheduling provides on-demand resource allocation, dynamic adaptation, and collaborative optimization across multiple terminals.

In energy efficiency optimization, AI can identify patterns of load variation by analyzing network load and traffic behavior over different time periods [59]. Using this insight, the system can selectively disable parts of the SFU infrastructure during idle hours, reducing standby power consumption and enabling coarse-grained power control. In addition, with enhanced service classification and QoS awareness, AI can apply differentiated power management strategies tailored to specific traffic types [60]. For example, when low-priority services are detected, the system can trigger module-level deep sleep or delayed wake-up mechanisms. Without compromising service quality, AI dynamically adjusts device operating rates and power consumption to enable fine-grained energy-saving optimization.

In troubleshooting, the large-scale deployment of SFUs poses new challenges to traditional operations and maintenance workflows. To address these challenges, the MFU can aggregate multi-source data reported by SFUs, including operational logs and link performance indicators. Based on this data, AI techniques can be employed to perform anomaly detection and root cause analysis [61]. By constructing diagnostic models that incorporate both statistical and temporal features, the system can accurately detect problems such as link disruption, rate instability, and optical power anomalies. It can also determine whether a fault is optical or wireless in origin. This AI-driven capability significantly enhances the O&M process by enabling precise fault localization, intelligent work order assignment and autonomous recovery, thereby improving network availability and maintenance efficiency.

### B. FTTR-Enabled AI Deployment

Traditional indoor access networks face considerable challenges in data availability, computational capacity, and model deployment when applying AI at the user side. In terms of data, AI model training requires rich, high-quality, and real-time traffic datasets. However, conventional architectures typically support only coarse-grained data collection, lack sufficient real-time capabilities, and pose potential risks of user privacy leakage [62]. In terms of computing power, edge devices in home and small-to-medium enterprise environments often have limited computing resources, making it difficult to deploy large-scale or highly complex AI models. In addition, computational demands in such scenarios are highly distributed and dynamic, resulting in low resource utilization and poor coordination between devices [63]. As for model deployment, most AI models are originally designed for cloud environments, which often suffer from high response latency and limited adaptability to local operating conditions, especially in scenarios that require real-time control [64].

FTTR provides a robust foundation for the application and continuous optimization of AI technologies. FTTR builds a distributed perception network, where each SFU node acts as an edge data collector, continuously capturing high-frequency dynamic information such as user traffic, channel conditions, and access behavior. These data are characterized by high temporal resolution, fine-grained spatial distribution, and strong service awareness, making them valuable inputs for training high-precision AI models. In addition, by leveraging the native collaborative architecture of FTTR, privacy-preserving mechanisms such as federated learning and differential privacy can be employed to enable secure data sharing across nodes [65].

In terms of computing power, both MFU and SFU nodes can be equipped with built-in or external AI processing units to form a room-scale edge computing network. SFUs, located in close proximity to end-user terminals, are responsible for low-latency data acquisition, preliminary processing, and lightweight local inference. MFUs, as near-end aggregation nodes, handle more complex tasks such as data fusion and core model inference. Additionally, the FTTR infrastructure provides an all-optical communication channel for data and computing exchange between edge nodes and remote computing centers (e.g., OLT), ensuring high throughput and low latency connectivity. The OLT can also be equipped with AI processing capabilities to support centralized inference, global model management, and long-term knowledge aggregation. This infrastructure plays a key role in model deployment, enabling edge-deployed AI models can efficiently support parameter updates, remote inference result retrieval, and collaborative data processing [66]. Based on the hierarchical architecture between edge (SFU) and near-end (MFU) nodes, AI workloads can be flexibly partitioned to support decentralized inference and cooperative task execution. Computational resources can be dynamically adjusted according to specific scenario requirements. For example, in smart home scenarios, SFU nodes can directly run lightweight models such as speech recognition or visual recognition to enable rapid local response while preserving user privacy. In enterprise or industrial monitoring environments, video and traffic data collected at SFUs can be rapidly aggregated over fiber to MFUs, where intensive data analysis and real-time alerting are performed.

*C. Integrated Sensing in FTTR*

The FTTR architecture provides an ideal platform for passive sensing integration by leveraging fiber-based backhaul/fronthaul and ubiquitous Wi-Fi coverage. Currently, Wi-Fi devices primarily sense environmental variations by extracting Channel State Information (CSI) from received signals. Compared with traditional indicators such as RSSI, CSI enables finer-grained sensing tasks and offers improved sensing accuracy [67]. However, CSI acquisition remains technically challenging due to factors such as sampling imbalance, transmitter-receiver synchronization, and signal distortion [68].

For sensing information acquisition, each SFU leverages Wi-Fi to obtain CSI, effectively addressing the limitations of CSI collection in traditional architectures. Specifically, SFUs sequentially broadcast control frames, while non-transmitting SFUs receive these frames and extract CSI measurements. This approach offers two key advantages. First, all SFUs adopt a unified hardware design and performance specification, which supports standardized CSI acquisition interfaces and reduces measurement distortion caused by hardware inconsistencies. Second, tight clock synchronization among SFUs can be achieved through the FTTR system, thereby eliminating phase and frequency offsets typically introduced by unsynchronized transceiver clocks. In addition to Wi-Fi-based sensing, the optical fiber infrastructure in FTTR can also function as a sensing medium for monitoring environmental parameters such as temperature, pressure, vibration, and light [69], [70]. This capability complements Wi-Fi sensing and further enhances overall network observability.

In the analysis and decision-making of sensing data, FTTR offers significant advantages over traditional access networks. Contemporary sensing data analysis often relies on machine learning and deep learning techniques, which impose substantial computational demands [71]. FTTR supports hierarchical processing by equipping SFUs, MFUs, and OLTs with heterogeneous computing capabilities. Specifically, SFUs are responsible for CSI acquisition and local preprocessing, MFUs execute lightweight models for real-time classification and prediction, and OLTs aggregate sensory inputs from across the network to perform model training, updating, and distribution. Moreover, FTTR can integrate sensing data from diverse sources, including optical fiber, Wi-Fi, mmWave radar, and user-end IoT devices such as cameras and microphones. MmWave sensing enhances spatial resolution for motion detection and gesture recognition, while IoT devices provide rich visual and acoustic inputs to support complex environmental perception. By integrating these heterogeneous sensing modalities, the FTTR architecture can construct a distributed, multimodal, and multilayer cooperative sensing network.

*D. Typical Applications of FTTR-based Sensing and Control*

Integrated sensing and communication provide strong support for the intelligent and adaptive management of FTTR networks. A representative application lies in energy efficiency optimization. By jointly sensing environmental conditions such as human activity and day-night transitions through Wi-Fi and optical fiber, the MFU can intelligently adjust device operating states [72]. For example, during nighttime idle periods, selected SFUs can reduce power consumption or enter sleep mode to conserve energy. In addition, user mobility and behavior inferred from sensing data enable optimized roaming handovers and dynamic bandwidth allocation, thereby reducing latency and improving resource utilization.

The FTTR-based sensing and communication framework supports a wide range of applications in smart homes, healthcare, and industrial environments. In intelligent care scenarios, the combination of Wi-Fi sensing and FTTR infrastructure enables posture recognition, fall detection, and non-invasive respiratory monitoring, which enhances user safety and health management [67]. In hospital environments, anomalies detected through Wi-Fi sensing can trigger real-time alerts from SFUs to centralized monitoring systems. The integration of mmWave technology further improves sensing resolution and enables accurate gesture recognition [39]. Combined with FTTR's edge computing capabilities and low-latency optical backhaul/fronthaul, this enables room-scale non-contact control and enhances user interaction in smart environments. In industrial and enterprise scenarios, FTTR enables real-time personnel localization, equipment monitoring, and environmental sensing. By integrating fiber-optic and wireless sensing technologies, the system supports accurate detection of environmental parameters such as temperature, humidity, vibration, and gas leakage, thereby improving safety and situational awareness.

V. CONCLUSION

This paper presents a comprehensive study of the FTTR system architecture and protocol stack, focusing on three key technical aspects: centralized scheduling and control, integrated management and maintenance, and green energy-saving mechanisms. A simplified FTTR architecture was proposed based on MAC and PHY layer convergence to enhance coordination and scheduling efficiency. To support unified network control, a remote management mechanism based on the OMCI was introduced, enabling end-to-end management across MFU and SFUs. Furthermore, a service-aware energy-saving framework was developed to dynamically adjust device operation based on traffic patterns and user behavior. The study also explored the integration of AI and passive sensing into FTTR networks, enabling intelligent resource scheduling, adaptive power control, and environment-aware optimization. These contributions provide technical support for the scalable, intelligent, and energy-efficient deployment of FTTR in future residential and enterprise networks.